\documentstyle[cite,psfig,epsf,epsfig]{kapproc}

\let\footnote\savefootnote

\normallatexbib

\begin{document}

\articletitle{Water at Positive and Negative\\ Pressures}

\author{H. E. STANLEY,$^1$ M. C. BARBOSA,$^{1,2}$ S. MOSSA,$^1$
P. A. NETZ,$^3$\\ F. SCIORTINO,$^4$ F. W. STARR,$^5$ and M. YAMADA$^1$}

\affil{$^1$Center for Polymer Studies and Department of Physics,
Boston University\\ Boston, MA 02215 USA\\
$^2$Instituto de F\'{\i}sica, Universidade Federal do Rio
Grande do Sul, Caixa Postal 15051\\ 91501-970, Porto Alegre,
RS, Brazil\\
$^3$Departamento de Qu\'{\i}mica, Universidade Luterana do
Brasil \\ 92420-280, Canoas, RS, Brazil \\
$^4$Dipartimento di Fisica Universit\`{a} di Roma La Sapienza,
Istituto Nazionale di Fisica\\ della Materia, and
INFM Center for Statistical Mechanics and Complexity,\\ Piazzale Aldo
Moro 2, 00185 Roma, Italy\\
$^5$Polymers Division and Center for Theoretical and
Computational Materials Science \\ National Institute of Standards and
Technology, Gaithersburg, MD 20899 USA\\}

\begin{abstract}

We review recent results of molecular dynamics simulations of two models
of liquid water, the extended simple point charge (SPC/E) and the
Mahoney-Jorgensen transferable intermolecular potential with five points
(TIP5P), which is closer to real water than previously-proposed
classical pairwise additive potentials.  Simulations of the TIP5P model
for a wide range of deeply supercooled states, including both positive
and negative pressures, reveal (i) the existence of a non-monotonic
``nose-shaped'' temperature of maximum density (TMD) line and a
non-reentrant spinodal, (ii) the presence of a low temperature phase
transition.  The TMD that changes slope from negative to positive as P
decreases and, notably, the point of crossover between the two behaviors
is located at ambient pressure (temperature $\approx 4 $ C , and density
$ \approx 1$ g/cm$^3$). Simulations on the dynamics of the SPC/E model
reveal (iii) the dynamics at negative pressure shows a minimum in the
diffusion constant $D$ when the density is decreased at constant
temperature, complementary to the known maximum of $D$ at higher
pressures, and (iv) the loci of minima of $D$ relative to the spinodal
shows that they are inside the thermodynamically metastable regions of
the phase-diagram.  These dynamical results reflect the initial
enhancement and subsequent breakdown of the tetrahedral structure and of
the hydrogen bond network as the density decreases.

\end{abstract}

\section{Introduction}

Water is an important liquid in nature, and is also fundamental in
chemical and technological applications. Although the individual water
molecule has a simple chemical structure, water is considered a complex
fluid because of its anomalous behavior
\cite{debe96,volga,ange82,four97,sta97a,lan82}.  It expands on freezing and,
at a pressure of 1 atm, the density has a maximum at
4$^\circ$C. Additionally, there is a minimum of the isothermal
compressibility at 46$^\circ$C and a minimum of the isobaric heat
capacity at 35$^\circ$C \cite{dou98}. These anomalies are linked with
the microscopic structure of liquid water, which can be regarded as a
transient gel---a highly associated liquid with strongly directional
hydrogen bonds \cite{geig79,Stanley80}. Each water molecule acts as both
a donor and an acceptor of bonds, generating a structure that is locally
ordered, similar to that of ice, but maintaining the long-range disorder
typical of liquids. Despite the extensive work that has been done on
water, many aspects of its behavior remain unexplained.

Several scenarios have been proposed to account for the the anomalous
behavior of the thermodynamic response functions on cooling, each
predicting a different behavior for the liquid spinodal, the line of the
limit of stability separating the region where liquid water is
metastable from the region where the liquid is unstable. (i) According
the stability-limit conjecture \cite{spe82b,spe87}, the pressure of the
spinodal line should decrease on cooling, become negative, and increase
again after passing through a minimum. It reenters the positive pressure
region of the phase diagram at a very low temperature, thereby giving
rise to a line of singularities in the positive pressure region, and
consequently the increase in the thermodynamic response functions on
cooling in the anomalous region is due to the proximity of this
reentrant spinodal. (ii) The critical point hypothesis
\cite{pses,mishima94,critical-point,ms98,tana96,t35}, proposes a new
critical point at the terminus of a first-order phase transition line
separating two liquid phases of different density. The anomalous
increases of the response functions, compressibility, specific heat, and
volume expansivity, is interpreted in terms of this critical
point. (iii) The singularity-free hypothesis
\cite{Stanley80,sas96,sas98} proposes that actually there is no
divergence close to the anomalous region; the response functions grow on
lowering temperature but remain finite, attaining maximum values. 

Water properties and anomalies can be strongly influenced by the
physical or chemical properties of the medium
\cite{debe96,ange82,four97,sta97a,net98a,koga98}. The effect not only of
applied pressure, but also of negative pressure (``stretching'') is
remarkable. The study of the behavior of this fluid under negative
pressures is relevant not only from the academic point of view, but also
for realistic systems. For example, negative pressures are observed
\cite{poc95}, and seem to play an important role in the mechanism of
water transport in plants. Therefore, properties that modify the
structure of water, especially if this modification is similar to the
effect of stretching (as is the case in some hydrogels
\cite{net98a}), also influence its dynamical behavior.

Dynamic properties, such as the diffusion constant, have been studied in
detail for water systems at atmospheric and at high positive pressures,
both experimentally \cite{jon76,pri87} as well as by computer
simulations \cite{ram87,sci91,bae94,har97,francesco,sta99e,scala}. The
increase of pressure increases the presence of defects and of
interstitial water molecules in the network \cite{sci91}. They disrupt
the tetrahedral local structure, weakening the hydrogen bonds, and thus
increasing the diffusion constant \cite{sta99e,scala}. However, a
further increase in the pressure leads to steric effects which works in
the direction of lowering the mobility. The interplay of these factors
leads to a maximum in the diffusion constant \cite{sta99e,scala} at some
high density $\rho_{\mbox{\scriptsize max}}$. Above this density (or
corresponding pressure), the diffusion of water is in some sense like
that of a normal liquid, controlled by hindrance, with the hydrogen
bonds playing a secondary role.  However, the behavior at very low
$\rho$ is less well understood.

\section{Location of the Spinodal at Positive and Negative Pressures}

Relatively few experimental works \cite{hen80,GDWA90} and simulations
\cite{pses,tana96,francesco,sta99e,err00,vai93} have been performed on
``stretched'' water.  In this negative pressure region of the phase
diagram the system is metastable, and becomes unstable beyond the
spinodal line, so locating the spinodal we can ensure that the simulated
state points lie in the metastable and not in the unstable
region. Moreover, the shape of the spinodal can test the stability-limit
conjecture against the critical point hypothesis and the
singularity-free interpretation, so we first discuss the density and
pressure of the spinodal, which we denote $\rho_{\mbox{\scriptsize
sp}}(T)$ and $P_{\mbox{\scriptsize sp}}(T)$, respectively.

Yamada and her coworkers \cite{RPD96} simulated a system of $N=343$
molecules interacting with the TIP5P potential~\cite{Mahoney00}.  TIP5P
is a five-site, rigid, non-polarizable water model, not unlike the ST2
model~\cite{Stillinger74}. The TIP5P potential accurately reproduces the
density anomaly at 1 atm and exhibits excellent structural properties
when compared with experimental data~\cite{Mahoney00,Head-Gordon}. The
TMD shows the correct pressure dependence, shifting to lower
temperatures as pressure is increased. Under ambient conditions, the
diffusion constant is close to the experimental value, with reasonable
temperature and pressure dependence away from ambient
conditions~\cite{Mahoney00}.
\begin{figure}[bht]
\begin{center}
\epsfig{file=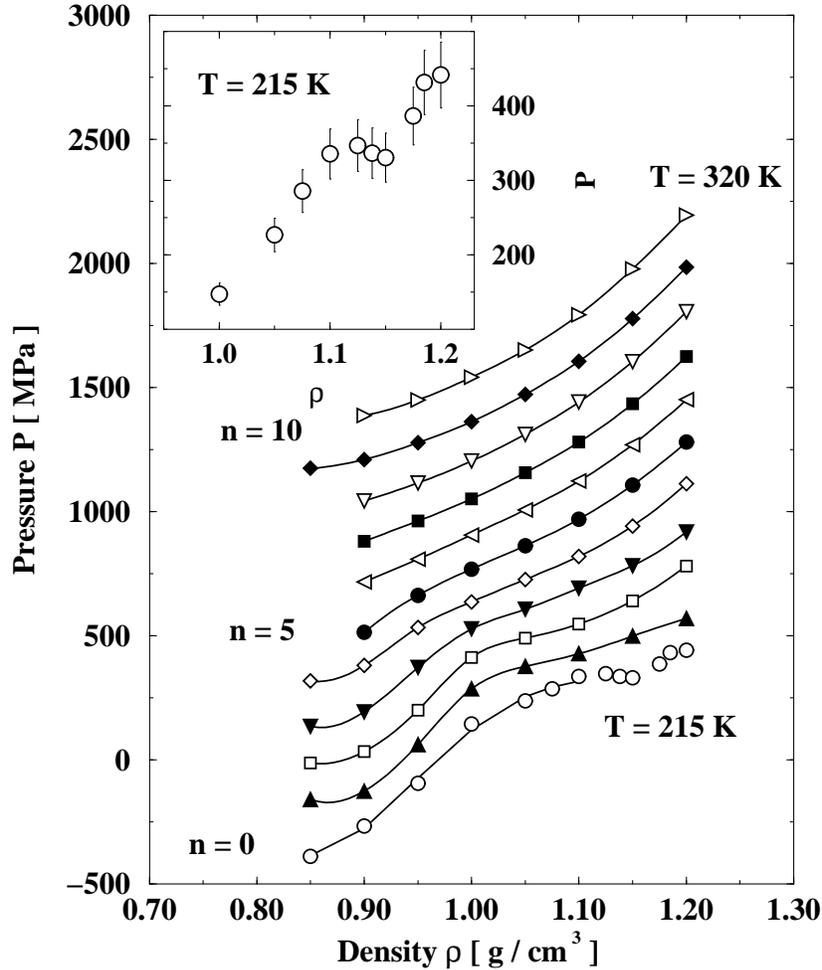,clip=,width=0.9\linewidth}
\end{center}
\caption{
\label{fig:isotherms}
Dependence on density of the pressure at all temperatures investigated
($T=215,220,230,240,250,260,270,280,290,300,320$ K, from bottom to top).
Each curve has been shifted by $n\times 150$ MPa to avoid overlaps.  An
inflection appears as T is decreased, transforming into a ``flat''
coexistence region at $T=215$ K, indicating the presence of a
liquid-liquid transition.  Inset: A detailed view of the $T=215$ K
isotherm. Adapted from \protect\cite{RPD96}.}
\end{figure}
Equilibration runs were performed at constant $T$.  After thermalization
at $T =320$~K the thermostat temperature was set to the temperature of
interest.  The system evolved for a time longer than the structural
relaxation time $\tau_\alpha$, defined as the time at which
$F_s(Q_0,\tau_\alpha)=1/e$, where $F_s(Q_0,t)$ is the self-intermediate
scattering function evaluated at $Q_0=18$ nm$^{-1}$, the location of the
first peak of the static structure factor.  In the time $\tau_\alpha$,
each molecule diffuses on average a distance of the order of the nearest
neighbor distance.  We use the final configuration of the equilibration
run to start a production run of length greater than several
$\tau_\alpha$ and then analyze the calculated trajectory.

\begin{figure}[bht]
\begin{center}
\epsfig{file=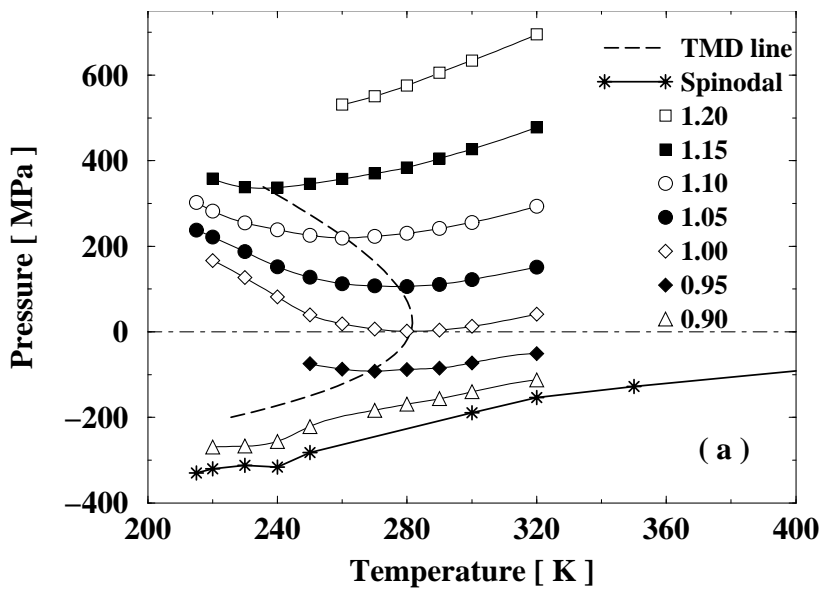,clip=,width=0.9\linewidth}
\end{center}
\caption{
\label{fig:phase_diagram} 
Pressure along seven isochores; the minima correspond to the
temperature of maximum density line (dashed line). Note the ``nose'' of
the TMD line at $T=4$ C. Stars denote the liquid spinodal line, which is
not reentrant, and terminates at the liquid-gas critical point. Adapted
from \protect\cite{RPD96}.}
\end{figure} 

Figure~\ref{fig:isotherms} shows results for pressure along
isotherms. At lower temperatures an inflection develops, which becomes a
``flat'' isotherm at the lowest temperature, $T=215$ K.  The presence of
a flat region indicates that a phase separation takes place; the
critical temperature is $T_{C'}=217 \pm 3$ K, the critical pressure is
$P_{C'}=340\pm$ 20 MPa, and the critical density $\rho_{C'}=1.13 \pm
0.04$ g/cm$^3$.

Figure~\ref{fig:phase_diagram}(a) plots the pressure along isochores.
The curves show minima as a function of temperature; the locus of the
minima is the TMD line, since $(\partial P/\partial T)_V=\alpha_P /K_T$.
Note that the pressure exhibits a minimum if the density passes through
a maximum ($\alpha_p=0$).  It is clear that, as in the case of ST2 water,
TIP5P water has a TMD that changes slope from negative to positive as P
decreases. Notably, the point of crossover between the two behaviors is
located at ambient pressure, $T\approx 4 $ C , and $\rho \approx 1$
g/cm$^{3}$.

Also plotted the spinodal line, obtained by fitting the isotherms (for
$T\ge 300 K$) of Fig.~\ref{fig:isotherms} to the form
$P(T,\rho)=P_s(T)+A\left[\rho-\rho_s(T)\right]^2$, where $P_s(T)$ and
$\rho_s(T)$ denote the pressure and density of the spinodal line.  This
functional form is the mean field prediction for $P(\rho)$ close to a
spinodal line.  For $T\le 250 K$, $P_s(T)$ is calculated by estimating
the location of the minimum of $P(\rho)$.  The results in
Fig.\ref{fig:phase_diagram} show that the liquid spinodal line is not
reentrant and does not intersect the TMD line.

\section{Dynamic Properties}

We next discuss results on the dynamics of stretched water recently
obtained by Netz and his collaborators \cite{NetzXX}. While there are a
large number of intermolecular potential functions used to simulate
water, each of which gives slightly different results, the overall
thermodynamics picture obtained from these models is generally very
similar. Since dynamic properties are particularly sensitive to the
potential choice, the extended simple point charge (SPC/E) potential is
used since it reproduces both the maximum in diffusivity under pressure
as well as the power-law behavior of dynamics properties on cooling.
For understanding the properties of water at negative pressure,
simulations are particularly important since experiments are very
difficult to perform in this region.

\begin{figure}[bht]
\begin{center}
\epsfig{file=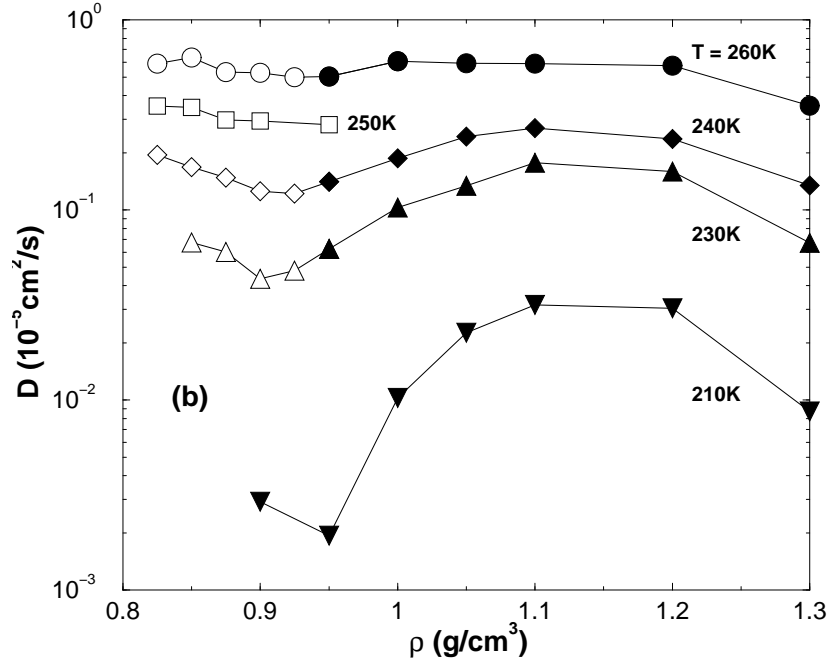,clip=,angle=-90,width=0.9\linewidth}
\end{center}
\caption{Dependence of the diffusion constant $D$ on $\rho$ along
isotherms (for $\rho \le 1.0$~g/cm$^3$).  Open symbols are the new
simulations we report, and filled symbols are from
Ref.~\protect\cite{sta99e}. Adapted from \protect\cite{NetzXX}.}
\label{f4}
\end{figure}

\begin{figure}[bht]
\begin{center}
\epsfig{file=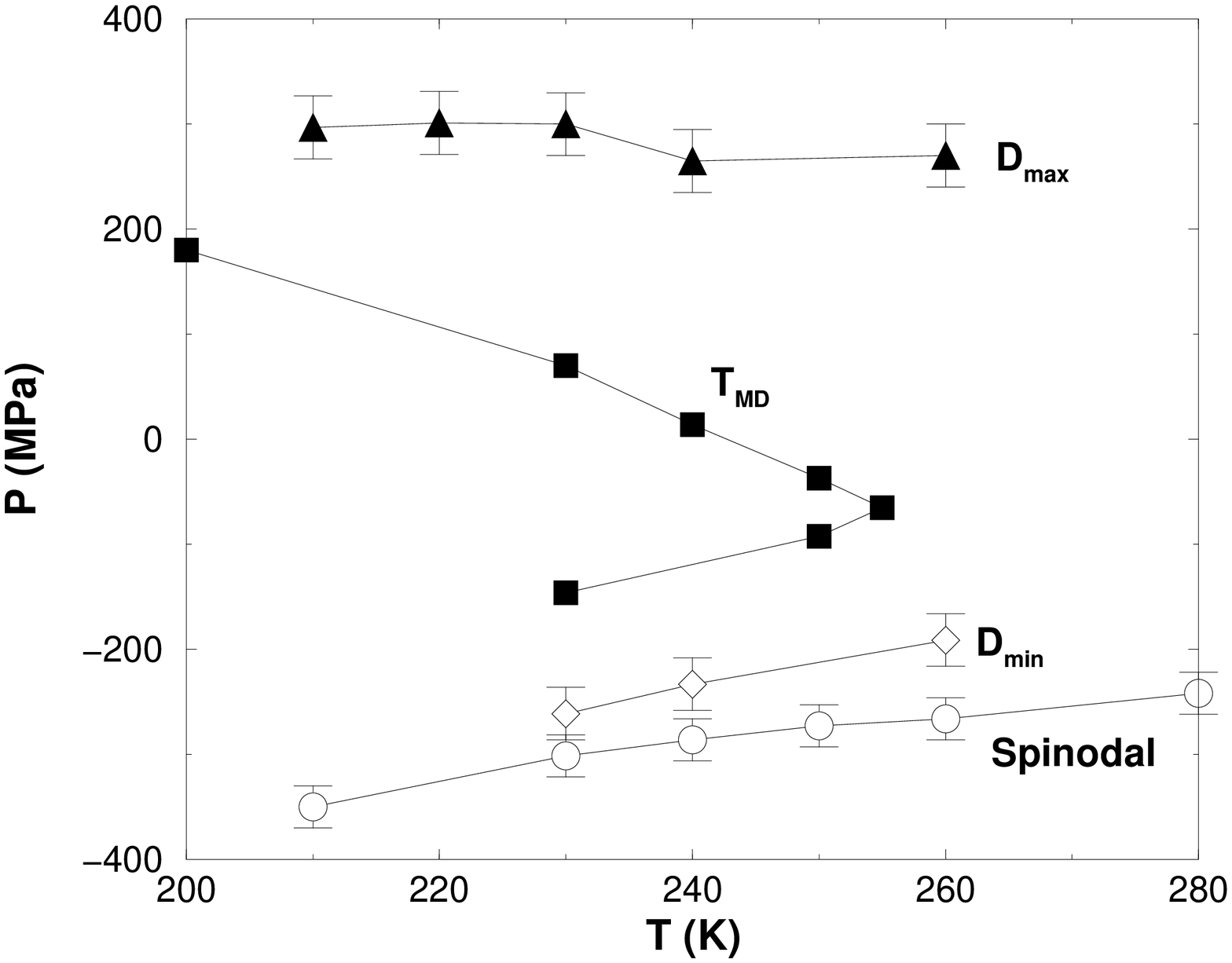,clip=,angle=-90,width=0.9\linewidth}
\end{center}
\caption{Relation of the loci of maxima and minima of $D$ with 
$T_{MD}$ and the spinodal.  Open symbols are from the present work, and
filled symbols are from Ref.~\cite{sta99e}. Adapted from
\protect\cite{NetzXX}.}
\label{loci}
\end{figure}

The effect of extreme conditions on the flow of the liquid is assessed by
calculating the diffusion constant $D$, defined by the asymptotic value
of the slope of the mean square displacement versus time. We show $D$
along isotherms in Fig.~3.  For $T\le260$~K, $D$ has a minimum
value at $\rho\approx 0.9$~g/cm$^3$, which becomes more pronounced at
lower $T$.  This behavior can be understood considering
the structural changes that occur with decreasing density. At low $T$,
the decreased density enhances the local tetrahedral ordering, which
leads to a decrease in $D$.  Further decreases in density reduces the
stability of the tetrahedral structure and causes an increase of $D$.

The location of the minimum is near the ice Ih density $\approx
0.915$~g/cm$^3$, which is the density where the perfect tetrahedral
order occurs.  The behavior of the minimum of $D$, $D_{\rm min}(T)$,
complements the known behavior of $D_{\rm max}(T)$ for the same model
\cite{sta99e,scala,err00}, where a maximum occurs due to breaking
hydrogen bonds at high pressure; the density of the $D_{\rm min}(T)$
increases slightly with increasing $T$, while the density of $D_{\rm
max}(T)$ decreases with increasing $T$ \cite{scala}.  This is expected,
since the range of densities where anomalous behavior occurs expands
with decreasing $T$.  We show the loci of $D_{\rm min}(T)$ and $D_{\rm
max}(T)$, along with the spinodal and locus of density maxima in
Fig.~4.

Below the spinodal, $D$ also increases, since the mobility of the gas is
larger than that of the liquid.  However, the simulations clearly show that
$D_{\rm min}$ for the liquid occurs prior to the onset of cavitation,
and so the location of $D_{\rm min}$ we estimate is not affected by
phase separation.  Recently, Ref.~\cite{err00} estimated the location of
$D_{\rm min}$ for the same model along several isotherms, and associated
$D_{\rm min}$ with a maximum in orientational order.

\section{Conclusions}

Water exhibits a very complex structure and its properties and anomalies
are strongly influenced by variations of pressure.  For high densities
($\rho>\rho_{\mbox{\scriptsize max}}$), water behaves as a normal liquid
and the decrease of $D$ with increasing pressure is governed by steric
effects. For $\rho_{\mbox{\scriptsize min}} < \rho <
\rho_{\mbox{\scriptsize max}}$, as the pressure is decreased, the
presence of defects and interstitial water decrease, the tetrahedral
structure dominates, with stronger hydrogen bonds. This process reaches
its maximum at $\rho=\rho_{\mbox{\scriptsize min}} \approx
\rho_{\mbox{\scriptsize ice}}$. Further stretching destabilizes the
hydrogen bond network, leading to an increase in mobility. The locus of
$D_{\rm min}$ roughly tracks the spinodal, not surprising since the same
breakdown of tetrahedral order that gives rise to $D_{\rm min}$ also
facilitates cavitation.

\acknowledgments

We thank D.~R.~Baker, S. V. Buldyrev, P.~Debenedetti, G.~Franzese,
W.~Kob, E.~La Nave, M.~Marquez and C.~Rebbi for useful discussions, and
NSF Grant CHE-0096892, the Conselho Nacional de Desenvolvimento
Cientifico e Technologico (CNPq), the Fundacao de Amparo a Pesquisa do
Rio Grande do Sul (Fapergs) for support.  MY thanks NSF Grant
GER-9452651 for support as a Graduate Research Trainee at the Boston
University Center for Computational Science, FS thanks MURST COFIN 2000
and INFM Iniziativa Calcolo Parallelo, and FWS thanks the National
Research Council.


\begin{chapthebibliography}{999}

\bibitem{debe96} For elementary introductions to recent work on liquid
water, the reader may wish to consult P. Ball, {\it Life's Matrix: A
Biography of Water\/} (Farrar Straus and Giroux, New York,
2000) or P. G. Debenedetti and H. E. Stanley, ``The Novel Physics of Water
at Low Temperatures'', {\it Physics Today\/} (submitted).

\bibitem{volga} V. Brazhkin. S. V. Buldyrev, V. N. Ryzhov, and
H. E. Stanley [eds], {\it New Kinds of Phase Transitions:
Transformations in Disordered Substances} Proc. NATO Advanced Research
Workshop, Volga River (Kluwer, Dordrecht, 2002).

\bibitem{ange82} O. Mishima and H. E. Stanley, {\it Nature\/} {\bf 396}, 329 (1998).

\bibitem{four97} M.-C. Bellissent-Funel, ed., {\it Hydration Processes in Biology:
Theoretical and Experimental Approaches\/} (IOS Press, Amsterdam, 1999).

\bibitem{sta97a} H. E. Stanley, S. V. Buldyrev, N. Giovambattista, E. La
Nave, A Scala, F. Sciortino, and F. W. Starr, [Proc. IUPAP Statphys21,
Cancun] {\it Physica A} {\bf 306}, 230--242 (2002).

\bibitem{lan82} S. V. Buldyrev, G. Franzese, N. Giovambattista,
G. Malescio, M. R. Sadr-Lahijany, A. Scala, A. Skibinsky, and
H. E. Stanley [Proc. International Conf. on Scattering Studies of
Mesoscopic Scale Structure and Dynamics in Soft Matter] {\it Physica A}
{\bf 304}, 23-42 (2002).


\bibitem{dou98} R. C. Dougherty and L. N. Howard, {\it J. Chem. Phys.} {\bf 109}, 7379
(1998).

\bibitem{geig79} A. Geiger, F. H. Stillinger, and A. Rahman, 
%``Aspects of the Percolation
%Process for Hydrogen Bond Networks in Water,'' 
{\it J. Chem. Phys.} {\bf 70}, 4185 (1979).

\bibitem{Stanley80} H. E. Stanley and J. Teixeira, 
%``Interpretation of The Unusual Behavior
%of H$_2$O and D$_2$O at Low Temperatures: Tests of a Percolation
%Model,'' 
{\it J. Chem. Phys.} {\bf 73}, 3404 (1980).

\bibitem{spe82b} R. J. Speedy,
%``Limiting Forms of the Thermodynamic Divergence at the
%Conjectured Stability Limits in Superheated and Supercooled Water,''
{\it J. Chem. Phys.} {\bf 86}, 982 (1982); {\it Ibid} {\bf 86}, 3002
(1992).

\bibitem{spe87} R. J. Speedy,
%``Thermodynamics Properties of Supercooled Water at 1 atm.,'' 
{\it J. Chem. Phys.} {\bf 91}, 3354 (1987).

\bibitem{pses} P.~H. Poole, F. Sciortino, U. Essmann, and H.~E. Stanley,
{\it Nature\/} {\bf 360}, 324 (1992); {\it Phys. Rev. E\/} {\bf 48},
3799 (1993); F. Sciortino, P. H. Poole, U. Essmann, and H. E. Stanley,
Ibid. {\bf 55}, 727 (1997); S. Harrington, R. Zhang, P. H. Poole,
F. Sciortino, and H. E. Stanley, {\it Phys. Rev. Lett.} {\bf 78}, 2409
(1997).

\bibitem{mishima94} O. Mishima, {\it J. Chem. Phys.} {\bf 100}, 5910
(1994).

\bibitem{critical-point} P. H. Poole, F. Sciortino, T. Grande,
H. E. Stanley and C. A. Angell, {\it Phys. Rev. Lett.} {\bf 73}, 1632
(1994); C. F. Tejero and M.  Baus, {\it Phys. Rev. E\/} {\bf 57}, 4821
(1998); T. M. Truskett, P. G. Debenedetti, S. Sastry, and S. Torquato,
{\it J. Chem. Phys.} {\bf 111} 2647 (1999).


\bibitem{ms98} M.-C. Bellissent-Funel, {\it Europhys. Lett.} {\bf 42},
161 (1998); O. Mishima and H.~E. Stanley, Nature {\bf 392}, 192 (1998).

\bibitem{tana96} H. Tanaka,
%``Phase Behavior of Supercooled Water: Reconciling a Critical
%Point of Amorphous Ices with Spinodal Instability,'' 
{\it J. Chem. Phys.} {\bf 105}, 5099 (1996).

\bibitem{t35} A. Scala, F. W. Starr, E. La Nave, H. E. Stanley and
F. Sciortino, {\it Phys. Rev. E} {\bf 62}, 8016 (2000).

\bibitem{sas96} S. Sastry, P. G. Debenedetti, F. Sciortino, and
H. E. Stanley,
%``Singularity-Free Interpretation of the Thermodynamics of Supercooled
%Water,'' 
{\it Phys. Rev. E\/} {\bf 53}, 6144 (1996).

\bibitem{sas98} L. P. N. Rebelo, P. G. Debenedetti, and S. Sastry,
%``Singularity-Free Interpretation of the Thermodynamics of Supercooled
%Water: II. Thermal and Volumetric Behavior,'' 
{\it J. Chem. Phys.} {\bf 109}, 626 (1998).

\bibitem{net98a} P. A. Netz and Th. Dorfm\"uller,
%``Computer Simulation Studies on the Polymer-Induced Modification of
%Water Properties in Polyacrylamide Hydrogels,'' 
{\it J. Phys. Chem. B\/} {\bf 102}, 4875 (1998).

\bibitem{koga98} K. Koga, X. C. Zeng, and H. Tanaka,
%``Effects of Confinement on the
%Phase Behavior of Supercooled Water,''
{\it Chem. Phys. Lett.} {\bf 285}, 278 (1998).

\bibitem{poc95} W. T. Pockman, J. S. Sperry, and J. W. O'Leary,
%``Sustained and Significant Negative Water Pressure in Xylem,'' 
{\it Nature\/} {\bf 378}, 715 (1995).

\bibitem{jon76} J. Jonas, T. DeFries, and D. J. Wilbur, 
%``Molecular Motions in Compressed Liquid Water,'' 
{\it J. Chem. Phys.}  {\bf 65}, 582 (1976).

\bibitem{pri87} F. X. Prielmeier, E. W. Lang, R. J. Speedy, and H.-D. L\"udemann,
%``Diffusion in Supercooled Water to 300 MPa,'' 
{\it Phys. Rev. Lett.}  {\bf 59}, 1128 (1987);
%``The Pressure Dependence of Self-Diffusion in Supercooled Water and
%Heavy Water,'' 
{\it Ber. Bunsenges. Phys. Chem.} {\bf 92}, 1111 (1988).

\bibitem{ram87} M. Rami Reddy and M. Berkovitz, 
%``Structure and Dynamics of High-Pressure TIP4P Water,'' 
{\it J. Chem. Phys.} {\bf 87}, 6682 (1987).

\bibitem{sci91} F. Sciortino, A. Geiger, and H. E. Stanley, 
%``Effect of Defects on Molecular Mobility in Liquid Water,'' 
{\it Nature\/} {\bf 354}, 218 (1991); {\it Ibid.}, 
%``Network Defects and Molecular Mobility in Liquid Water,''
{\it J. Chem. Phys.} {\bf 96}, 3857 (1992).

\bibitem{bae94} N. Giovambattista, F. W. Starr, F. Sciortino,
S. V. Buldyrev, and H. E. Stanley, {\it Phys. Rev. E} {\bf 65}, 041502-1
-- 041502-6 (2002) cond-mat/0201028.

\bibitem{har97} E. La Nave, A. Scala, F. W. Starr, H. E. Stanley and
F. Sciortino, {\it Phys. Rev. E} {\bf 64}, 036102-1 -- 036102-10 (2001);
E. La Nave, H. E. Stanley and F. Sciortino, {\it Phys.~Rev.~Letters}
{\bf 88}, 035501-1 to 035501-4 (2002) cond-mat/0108546.

\bibitem{francesco} P. Gallo, F. Sciortino, P. Tartaglia, and
S.-H. Chen, Phys. Rev. Lett. {\bf 76}, 2730 (1996).

\bibitem{sta99e} F. W. Starr, F. Sciortino, and H. E. Stanley, 
%``Dynamics of Simulated Water under Pressure,'' 
{\it Phys. Rev. E\/} {\bf 60}, 6757 (1999); F. W. Starr,
S. T. Harrington, F.~Sciortino, and H. E. Stanley, {\em
Phys. Rev. Lett.}, {\bf 82}, 3629, (1999).

\bibitem{scala} A. Scala, F. W. Starr, E. La Nave, F. Sciortino and
H. E. Stanley, {\it Nature\/} {\bf 406}, 166 (2000).

\bibitem{hen80} S. J. Henderson and R. J. Speedy, 
%``A Berthelot-Bourdon Tube Method for Studying Water under Tension,''
{\it J. Phys. E: Scientific Instrumentation\/} {\bf 13}, 778 (1980).

\bibitem{GDWA90} J. L. Green, D. J. Durben, G. H. Wolf, and
C. A. Angell, {\it Science\/} {\bf 249}, R649 (1990).

\bibitem{vai93} I. I. Vaisman, L. Perera, and M. L. Berkovitz, 
%``Mobility of Stretched Water,'' 
{\it J. Chem. Phys.} {\bf 98}, 9859 (1993).

\bibitem{err00} J. R. Errington and P. G. Debenedetti,
{\it Nature\/} {\bf 409}, 318  (2001).

\bibitem{RPD96} M. Yamada, S. Mossa, H. E. Stanley, F. Sciortino,
{\it Phys.~Rev.~Letters\/} {\bf 88},  195701 (2002); cond-mat/0202094
%
\bibitem{Mahoney00}
M.~W.~Mahoney and W.~L.~Jorgensen, J. Chem. Phys. {\bf 112}, 8910
(2000);  {\it Ibid.} {\bf 114}, 363 (2001). 
%
\bibitem{Stillinger74} F.~H.~Stillinger and A.~Rahman, {\it
J. Chem. Phys.} {\bf 60}, 1545 (1974).
%
\bibitem{Head-Gordon} J.~M.~Sorenson, G.~Hura, R.~M.~Glaeser, and
T.~Head-Gordon, {\it J. Chem. Phys.}  {\bf 113}, 9149 (2000).

\bibitem{NetzXX} P. A. Netz, F. W. Starr, H. E. Stanley, and
M. C. Barbosa, {\it J. Chem. Phys.} {\bf 115}, 344--348 (2001);
cond-mat/0102196; P. A. Netz, F. W. Starr, H. E. Stanley, and
M. C. Barbosa, cond-mat/0201130; P. A. Netz, F. Starr, M. C. Barbosa,
H. E. Stanley, cond-mat/0201138.


%\end{thebibliography}

%=============================================================
%		*********************************** 
%=============================================================
\end{chapthebibliography}
\end{document}